\documentstyle[sprocl]{article}

\input{psfig}
\newcommand{\zp}[3]{{\sl Z. Phys.} {\bf #1} (19#2) #3}

\newcommand{\pl}[3]{{\sl Phys. Lett.} {\bf #1} (19#2) #3}

\newcommand{\prl}[3]{{\sl Phys. Rev. Lett.} {\bf #1} (19#2) #3}

\newcommand{\desy}[1]{{\sl DESY-Report~}{#1}}

\def\as{\alpha_s}
\def\sq{\tilde{q}}
\def\sqb{\bar{\tilde{q}}}

\def\gl{\tilde{g}}
\def\ppb{p\bar{p}}
\def\ms{m_{\tilde{q}}}
\def\mg{m_{\tilde{g}}}
\def\eps{\varepsilon}

\def\MS{$\overline{MS}$~}

\def\be{\begin{equation}}
\def\ee{\end{equation}}
\def\bea{\begin{eqnarray}}
\def\eea{\end{eqnarray}}

\begin{document}

\title{CROSS-SECTIONS FOR SQUARK AND GLUINO PRODUCTION AT HADRON
  COLLIDERS} 

\author{ R. H\"OPKER}

\address{Deutsches Elektronen-Synchrotron DESY, D-22603 Hamburg, Germany}

\author{ W. BEENAKKER}

\address{Instituut-Lorentz, University of Leiden, The Netherlands}

\maketitle\abstracts{ We present the cross-sections for the
  hadroproduction of squarks and gluinos in next-to-leading order of
  supersymmetric QCD. The four possible final states
  squark-antisquark, squark-squark, gluino-gluino and squark-gluino
  are analysed for the hadron colliders Tevatron and LHC. The
  dependence of the cross-sections on the renormalization and
  factorization scale is reduced significantly. The shape of the
  transverse-momentum and rapidity distributions remains nearly
  unchanged when the next-to-leading order SUSY-QCD contributions are
  included.  The size of the corrections at the central scale, given
  by the average mass of the produced particles, varies between $+5\%$
  and $+90\%$.}

\section{Introduction}
The colored squarks ($\sq_L,\sq_R$) and gluinos ($\gl$), the
supersymmetric partners of the quarks ($q$) and gluons ($g$), can be
searched for most efficiently at high-energy hadron colliders. As
R-parity is conserved in the QCD sector of the Minimal Supersymmetric
Standard Model (MSSM), these particles are always produced in pairs.
At the moment they can be searched for at the Fermilab Tevatron, a
$\ppb$ collider with a centre-of-mass energy of $1.8$TeV
[Ref.\,\cite{cdf,d0}].  In the future the CERN Large Hadron Collider
(LHC), the $pp$ collider with an envisaged centre-of-mass energy of
$14$TeV, will allow to cover mass values up to 1--2\,TeV
[Ref.\,\cite{lhc}].

So far most of the experimental analyses have been based on the
lowest-order (LO) production cross-sections. For obtaining
adequate theoretical predictions the LO cross-sections are in general
not sufficient. The most important arguments in favor of an analysis
that takes into account the next-to-leading-order (NLO) SUSY-QCD
corrections are:
\begin{itemize}
\item The LO cross-sections have a strong dependence on the {\it a
    priori} unknown renormalization scale $Q_R$ and the factorization
  scale $Q_F$.  Consequently, the theoretical predictions have in
  general an uncertainty that is almost as large as the cross-section
  itself. By implementing the NLO corrections a substantial reduction
  of the scale dependence is expected.
\item From experience with similar processes (e.g.~hadroproduction of
  top quarks), the NLO QCD corrections are expected to be sizeable.
\item An enhancement of the cross-section would lead to a higher value
  for the lower mass bounds for squarks and gluinos.
\item In case of discovery of squarks and gluinos, a precise knowledge
  of the total cross-sections would be mandatory for the determination
  of the masses of the particles.
\end{itemize}

Here we report on the calculations of NLO SUSY-QCD corrections to the
production of squarks and gluinos in $p\bar{p}/pp$ collisions, based
on the studies presented in Ref.\,\cite{BHSZprod}.  For a more
comprehensive report on this topic we refer to
Ref.\,\cite{BHSZreport}.

\section{Technical set-up}

\begin{figure}[h]
  \begin{center}
  \psfig{file=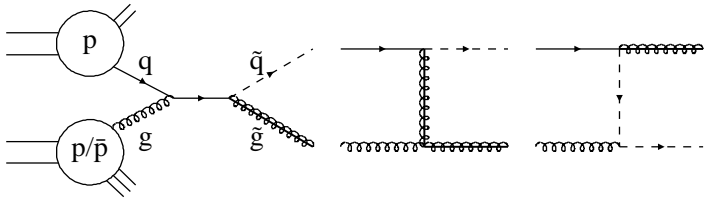,width=12cm}
  \end{center}
  \caption[]{Generic Feynman diagrams for squark-gluino production 
    in $\ppb/pp$ collisions.}
  \label{fig:hadronprod}
\end{figure}

We consider the following hadronic production processes:
\begin{equation}
 \ppb/pp \to \sq\sqb,\,\sq\sq,\,\gl\gl,\,\sq\gl \qquad (\sq\neq\tilde{t}),
\end{equation}
where the chiralities and flavors of the squarks
(e.g.~$\tilde{u}_L,\,\tilde{d}_R$) as well as the charge-conjugate
final states (e.g.~$\sqb\sqb$) are implicitly summed over. They are
exemplified in Fig.\ref{fig:hadronprod} for squark-gluino production.

In analogy to the experimental analysis, we exclude top-squarks from
the final state and take all produced squarks to be mass degenerate. A
study of the production of pairs of top-squarks is in progress,
including the (model-dependent) mixing effects in the squark sector.
At the partonic level many subprocesses contribute at LO and NLO,
corresponding to different flavors/chiralities of the squarks and
different initial-state partons. The initial-state partons are made up
of the massless gluons and the five light quark flavors ($n_f=5$),
which are considered to be massless as well.

\begin{figure}[h]
  \begin{center}
    \psfig{file=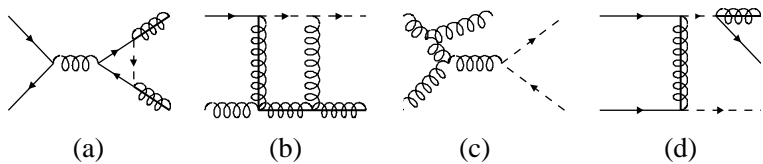,width=12cm}
  \end{center}
  \caption[]{Generic virtual corrections [(a): vertex contributions, (b): box
    contributions] and real radiation [(c): gluon radiation, (d):
    quark radiation].}
  \label{fig:hadronvirt}
\end{figure}

The NLO SUSY-QCD corrections include the virtual corrections
(consisting of self-energy corrections, vertex corrections
Fig.\ref{fig:hadronvirt}(a), and box diagrams (b)), real-gluon
radiation (c), and the radiation of a massless quark (d).

For the particles inside the loops we use the complete supersymmetric
QCD spectrum, i.e., gluons, gluinos, all quarks, and all squarks. We
have excluded the top-squarks from the final states. In order to have
a consistent NLO calculation, however, we have to take the top-squarks
into account inside loops. For the sake of simplicity we take them
mass-degenerate with the other squarks. For the top quark we use
$m_t=175$~GeV. Consequently the final results will depend on two free
parameters: the squark mass $\ms$ and the gluino mass $\mg$.

The divergences appearing in the NLO corrections are regularized by
performing the calculations in $n=4-2\eps$ dimensions. These
divergences consist of ultraviolet (UV), infra-red (IR), and collinear
divergences, and show up as poles of the form $\eps^{-i}$~($i=1,2$).
For the treatment of the $\gamma_5$ Dirac matrix, entering through the
quark--squark--gluino Yukawa couplings, we use the `naive' scheme.
This involves a $\gamma_5$ that anticommutes with the other gamma
matrices. This is a legitimate scheme at the one-loop level for
anomaly-free theories.

The UV divergences can be removed by renormalizing the coupling
constants and the masses of the massive particles. In the case of the
mass renormalization we have used the on-shell scheme and for the
renormalization of the QCD coupling constant the modified Minimal
Subtraction (\MS) scheme. To preserve the supersymmetric relation
between the gauge coupling (e.g. in the quark-quark-gluon vertex) and
the Yukawa coupling (e.g. in the quark-squark-gluino vertex) we have
to add a finite renormalization to the Yukawa coupling
[Ref.\,\cite{yukawa}].  The heavy particles (top, squarks and gluinos)
are decoupled from the running of the strong coupling.

After carrying out this renormalization program, the cross-sections
are UV finite. Nevertheless there are still divergences left over. The
IR divergences cancel in the sum of virtual corrections and
soft-gluon radiation. In order to separate soft from hard radiation a
cut-off $\Delta$ is introduced in the invariant mass corresponding to
the radiated gluon and one of the produced massive particles. If soft
and hard contributions are added up, any $\Delta$ dependence
disappears from the cross-sections for $\Delta \to 0$. The remaining
collinear singularities, finally, can be absorbed into the
renormalization of the parton densities, carried out in the \MS
mass-factorization scheme.

If the gluinos are lighter than the squarks, gluinos can also be decay
products of on-shell squarks, $\sq \to \gl q$ [see
e.g.\,Fig.\ref{fig:hadronvirt}(d)]. However, these situations are
already accounted for by the LO cross-section (e.g. $\sq\sq$
production) and the subsequent decay. In order to avoid double
counting, we restrict ourselves to irreducible final states in which
gluinos do not evolve from on-shell squarks. For the wedge $\mg > \ms$
we disregard, in the same sense as above, the decay of the gluinos to
squarks.

\section{Results}

The hadronic cross-sections are obtained by the convolution of the
parton densities with the corresponding partonic cross-sections.  When
discussing LO and NLO results, we calculate all quantities
[$\as(Q_R^2)$, the parton densities, and the partonic cross-sections]
in LO and NLO, respectively. We exemplify our results by gluino-pair
production at the Tevatron and squark-gluino production at the LHC.
The qualitative behavior of the other processes is quite similar.

\begin{figure}[h]
  \begin{center}
    \psfig{file=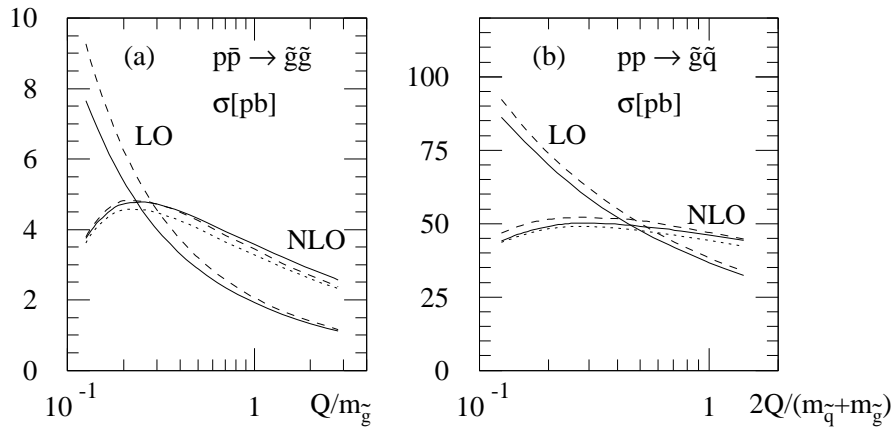,width=12cm}
  \end{center}
  \caption[]{Scale dependence of the total hadronic cross-sections for
    (a): gluino-pair production at the Tevatron ($\ms=280$GeV,
    $\mg=200$GeV, $\sqrt{S}=1.8$TeV), and (b): squark-gluino production
    at the LHC ($\ms=600$GeV, $\mg=500$GeV, $\sqrt{S}=14$TeV). Parton
    densities: GRV94 (solid), CTEQ3 (dashed), and MRSA' (dotted).}
  \label{fig:sigscale}
\end{figure}

As shown in Fig.\ref{fig:sigscale}, we find that the theoretical
predictions for the production cross-sections are nicely stabilized by
taking into account the NLO SUSY-QCD corrections.  In all processes,
for both the Tevatron and the LHC, it is observed that the dependence
on the renormalization/factorization scale $Q$ ($Q=Q_R=Q_F$) is quite
steep and monotonic in LO, whereas the $Q$ dependence is reduced
significantly in NLO. Even a broad maximum develops at scales near one
third of the average mass of the final-state particles. The variation
of the cross-sections as a result of different NLO parametrizations of
the parton densities is $\leq 10\%$ at the Tevatron and $\leq 13\%$ at
the LHC, where the gluon densities are more important.

\begin{figure}[h]
  \begin{center}
    \psfig{file=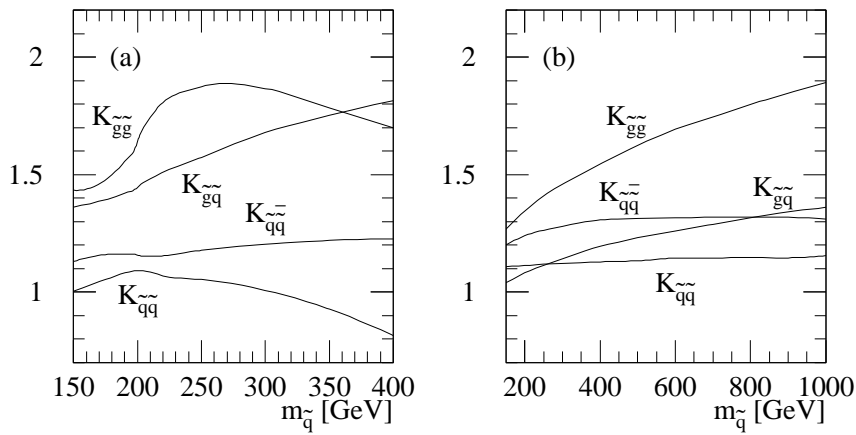,width=12cm}
  \end{center}
  \caption[]{$K$-factors ($K=\sigma_{NLO}/\sigma_{LO}$) for the four
    possible final states at (a): the Tevatron ($\mg=200$GeV), and (b):
    the LHC ($\ms/\mg=1.2$). Parton densities: GRV94.}
  \label{fig:sigkfac}
\end{figure}

From now on we use GRV94 parton densities and conservatively take as
default scale $Q$ the average mass of the produced particles. The
$K$-factors, $K=\sigma_{NLO}/\sigma_{LO}$ , depend strongly on the
process. This is exemplified in Fig.\ref{fig:sigkfac}(a) for the
Tevatron and in (b) for the LHC.  For both collider types the NLO
corrections are positive and large (up to $+90\%$) for the dominant
production cross-sections, involving at least one gluino in the final
state. The corresponding $K$-factors also exhibit a sizeable
dependence on the squark and gluino masses (especially the $\ms$
dependence of $K_{\gl\gl}$ is large). The NLO corrections for squark
final states are moderate ($\leq +30\%$). In case of discovery of
squarks and gluinos, the inclusion of the NLO SUSY-QCD corrections is
needed for an accurate determination of the cross-sections and the
masses.

\begin{figure}[h]
  \begin{center}
    \psfig{file=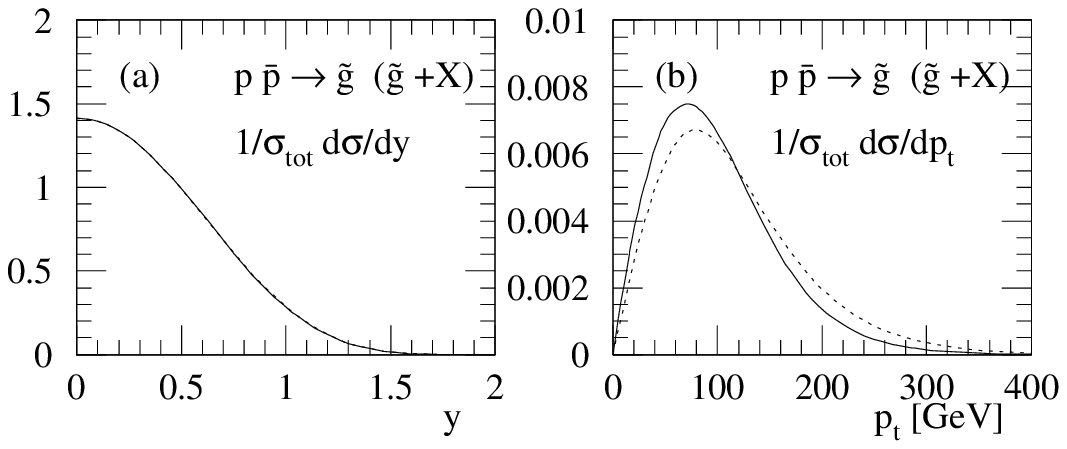,width=12cm}
  \end{center}
  \caption[]{Normalized differential cross-sections for gluino-pair
    production at the Tevatron ($\ms=280$GeV, $\mg=200$GeV) with
    respect to (a): the rapidity of one gluino, and (b): the
    transverse momentum of one gluino. NLO (solid) and LO (dotted).
    Parton densities: GRV94.}
  \label{fig:sigdiffgg}
\end{figure}

\begin{figure}[h]
  \begin{center}
    \psfig{file=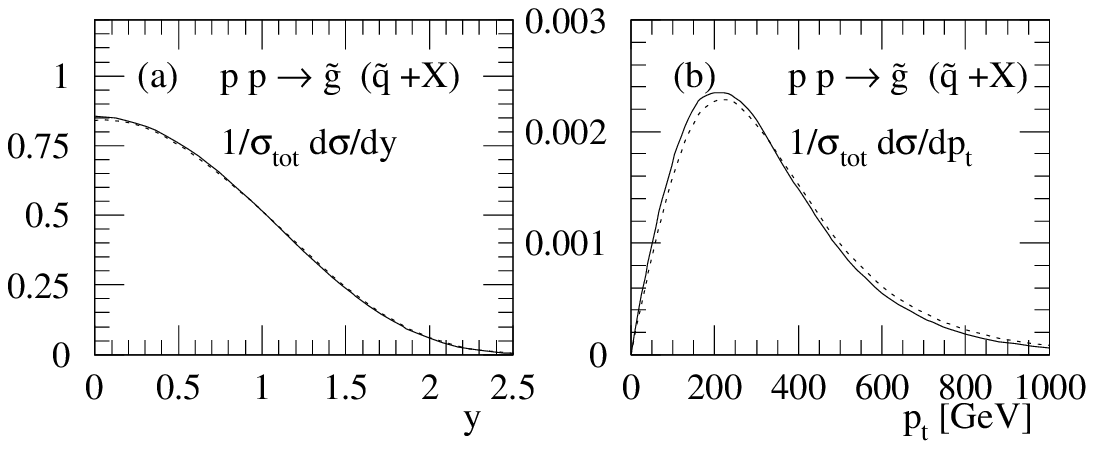,width=12cm}
  \end{center}
  \caption[]{Normalized differential cross-sections for squark-gluino
    production at the LHC ($\ms=600$GeV, $\mg=500$GeV) with respect to
    (a): the rapidity of the gluino, and b): the transverse momentum
    of the gluino. NLO (solid) and LO (dotted). Parton densities:
    GRV94.}
  \label{fig:sigdiffsg}
\end{figure}

Apart from the total cross-sections, also distributions with respect
to the rapidity ($y$) and transverse momentum ($p_t$) of one of the
outgoing massive particles can be studied. The differential
cross-sections are shown for gluino-pairs at the Tevatron in
Fig.\ref{fig:sigdiffgg} and for squark-gluino at the LHC in
Fig.\ref{fig:sigdiffsg}. They are normalized to the total
cross-section in the corresponding order.  The $K$-factors for these
distributions are independent of $y$ for all practical purposes and
hardly depend on $p_t$, where the NLO corrections render the
distributions somewhat softer.  Consequently, multiplication of the LO
distributions with the above-defined $K$-factors for the total
cross-sections approximates the full NLO results quite well.

Comparison of the NLO cross-sections with the cross-sections used in
the experimental Tevatron analyses (LO, EHLQ parton densities, and a
scale $Q$ that equals the partonic centre-of-mass energy) reveals that
the NLO corrections raise the lower mass bounds for squarks and
gluinos by $+10$GeV to $+30$GeV. For the LHC the shift in the mass
values between LO and NLO cross-sections amounts to $+10$GeV to
$+50$GeV.

\section*{Acknowledgments}
W. Beenakker is supported by a fellowship of the Royal Dutch Academy
of Arts and Sciences.

\end{document}